\documentclass[12pt]{article}
\usepackage{epsfig}
\usepackage{graphicx}
\usepackage{citesort}
\usepackage{amsmath,amssymb,amsfonts}

\usepackage[dvips,a4paper,textwidth=15.0cm,textheight=21.0cm]{geometry}
\newcommand{\beq}{\begin{equation}}
\newcommand{\eeq}{\end{equation}}
\newcommand{\bea}{\begin{eqnarray}}
\newcommand{\eea}{\end{eqnarray}}

\begin{document}
\title{\bf Quark number susceptibilities: \\lattice QCD versus PNJL model
\footnote{Work supported in part by BMBF, GSI and INFN}}
\author{
C. Ratti$^{a}$, S. R\"o{\ss}ner$^{b}$ and W. Weise$^{b}$\\
\\{\small $^a$ ECT$^*$, I-38050 Villazzano (Trento) Italy and INFN, 
Gruppo Collegato di Trento, }\\
{\small via Sommarive, I-38050 Povo (Trento) Italy}\\
{\small $^b$ Physik-Department, Technische Universit\"at M\"unchen, D-85747 
Garching, Germany}
}
\date{\today}
\maketitle
\begin{abstract}
Quark number susceptibilities at finite quark chemical potential are 
investigated in the framework of the Polyakov-loop-extended Nambu Jona-Lasinio
(PNJL) model. A detailed comparison is performed between the available lattice 
data, extrapolated using a Taylor expansion around vanishing chemical 
potential, 
and PNJL results consistently obtained from a Taylor series truncated at the 
same order. The validity of the Taylor expansion is then examined through a 
comparison between the full and truncated PNJL model calculations.
\end{abstract}

Investigations of QCD thermodynamics using lattice simulations predict a
transition from the hadronic to the quark-gluon phase
around a critical temperature of 0.2 GeV, at vanishing quark chemical potential
$\mu_q$.
This transition is signalled by a steep rise in energy density, pressure 
and entropy density as functions of the temperature, 
and known to be a crossover \cite{Fodor1} when quarks are included.

Based on model calculations
\cite{Asakawa:1989bq} as well as 
lattice QCD simulations~\cite{Fodor:2001au}, 
the existence
of a critical endpoint in the phase diagram is suggested. The precise location 
of this point is still under debate. It can possibily be identified
by observables that are sensitive to singular parts of the free energy
\cite{Ejiri:2005wq}, such as quark number susceptibilities.
Such susceptibilities have been studied both in lattice calculations
\cite{Allton:2003vx,Allton:2005gk}, and using phenomenological models
\cite{Liao:2005pa,Schaefer:2006ds,Majumder:2006nq,Sasaki:2006ws}. 
Both diagonal
and off-diagonal susceptibilities can serve as useful diagnostics
to elucidate the nature of strongly interacting matter~\cite{Koch:2005vg}.

The first lattice results for quark number 
susceptibilities at finite chemical potential have been obtained in
\cite{Allton:2003vx} through the Taylor expansion method. An
expansion of the pressure in powers of $\mu_q/T$ around $\mu_q=0$ was performed
up to fourth order in $\mu_q/T$. Quark number 
susceptibilities
were then obtained as second derivatives of the pressure with respect to 
the chemical potential. The presence of a peak in these observables at large 
chemical potentials has been interpreted as a signal of 
growing fluctuations in the baryon density, as the critical endpoint is 
approached. Improvements were implemented in
\cite{Allton:2005gk} by expanding the pressure up to sixth order in $\mu_q/T$.
In this case, the peak was shifted to a 
smaller temperature, and a dip was observed at $T\sim1.05~T_c$ which,
together with the increased error bars in the extrapolated lattice data, made 
the presence of the peak less convincing.

In the present paper we investigate quark number susceptibilities within
the Polyakov-loop-extended Nambu Jona-Lasinio (PNJL) model
\cite{Fukushima:2003fw,Meisinger:1995,Ratti:2005jh,Megias:2004hj} which
has recently been used successfully in comparisons with a variety of lattice
QCD data
\cite{Ghosh:2006qh,Hansen:2006ee,Ratti:2006wg,Mukherjee:2006hq,Rossner:2006xn,Zhang:2006gu}.
In this model, quarks propagate in a temporal background gauge field representing
Polyakov loop dynamics, while developing at the same time a dynamical mass
through their coupling to the chiral condensate. Here we present results for 
the
quark number susceptibilities at finite chemical potential. A comparison is 
first performed with truncated Taylor expansions in $\mu_q/T$ derived from
lattice QCD. Then the convergence properties of the Taylor series are examined
by comparison with the full, non-truncated PNJL result.

The Euclidean action of the two-flavor PNJL model is
\beq 
{\cal S}_E(\psi, \psi^\dagger, \phi)= \int _0^{\beta=1/T} d\tau\int_V d^3x \left[\psi^\dagger\,\partial_\tau\,\psi + {\cal H}(\psi, \psi^\dagger, \phi)\right] 
-\frac{V}{T}\,\mathcal{U}(\phi,T).
\label{action}
\eeq 
Here $\mathcal{H}$ is the fermionic Hamiltonian density given
by:
\beq
{\cal H} = -i\psi^\dagger\,(\vec{\alpha}\cdot \vec{\nabla}+\gamma_4\,m_0 -\phi)\,\psi + {\cal V}(\psi, \psi^\dagger)~,
\label{H}
\eeq
where $\psi$ is the $N_f=2$ doublet quark field, $\vec{\alpha} = \gamma_0\,
\vec{\gamma}$ and $\gamma_4 = i\gamma_0$ in terms of the standard 
Dirac $\gamma$ matrices and $m_0 = diag(m_u,m_d)$ is the quark mass matrix. 
$\mathcal{V}(\psi, \psi^\dagger)$ is an $SU(2)\times SU(2)$ invariant 
four-fermion interaction acting in the pseudoscalar-isovector/scalar-isoscalar 
quark-antiquark channel\footnote{The scalar diquark interaction that has been 
considered in \cite{Rossner:2006xn} is not relevant in the present discussion 
since the range of chemical potentials explored here is lower than the critical
$\mu_q$ for diquark condensation.}
\bea
\mathcal{V}\left(\psi,\psi^\dagger\right)= -\frac
{G}{2}\left[\left(\bar{\psi}\psi\right)^2+\left(\bar{\psi}\,i\gamma_5
\vec{\tau}\,\psi
\right)^2\right].
\label{V}
\eea

The quarks move in a background color gauge field $\phi \equiv A_4 = iA_0$, 
where $A_0 = \delta_{\mu 0}\,g{\cal A}^\mu_a\,t^a$ with the $SU(3)_c$ gauge 
fields ${\cal A}^\mu_a$ and the generators $t^a = \lambda^a/2$. The matrix 
valued, constant field $\phi$ relates to the (traced) Polyakov loop as follows:
\beq
\Phi=\frac{1}{N_c}\mathrm{Tr}\left[\mathcal{P}\exp\left(i\int_{0}^{\beta}
d\tau A_4\right)\right]=\frac{1}{ 3}\mathrm{Tr}\,e^{i\phi/T}~,
\eeq
In a convenient gauge (the so-called Polyakov gauge), one can choose a diagonal
representation for the matrix $\phi$,
\beq
\phi = \phi_3\,\lambda_3 +  \phi_8\,\lambda_8~,
\eeq
which leaves only two independent variables, $\phi_3$ and $\phi_8$.

The term $-(V/T)~{\cal U}$ of the 
action (\ref{action}) involves the effective potential   
$\mathcal{U}(\Phi,\Phi^*,T)$ which controls the thermodynamics of the Polyakov 
loop.
An improved form for this effective potential was 
introduced in our previous work~\cite{Rossner:2006xn}.
It involves the logarithm of $J(\Phi)$, the Jacobi determinant which results 
from integrating out six non-diagonal $SU(3)$ generators while keeping the two 
diagonal ones, $\phi_{3,8}$, to represent $\Phi$:
\beq
\frac{{\cal U}(\Phi,
\Phi^*,T )}{T^4}=-\frac{1}{2}a(T)\,\Phi^*\Phi
+b(T)\,\ln\left[1-6\,\Phi^*\Phi+4\left({\Phi^*}^3+\Phi^3\right)
-3\left(\Phi^*\Phi\right)^2\right]
\label{u1}
\eeq
with 
\beq
a(T)=a_0+a_1\left(\frac{T_0}{T}\right)
+a_2\left(\frac{T_0}{T}\right)^2,~~~~~~b(T)=b_3\left(\frac{T_0}{T}
\right)^3.
\label{u2}
\eeq
The logarithmic divergence of $\mathcal{U}(\Phi,\Phi^*,T)$ as 
$\Phi,~\Phi^*\rightarrow 1$ automatically limits the Polyakov loop $\Phi$ to be
always smaller than 1, reaching this value asymptotically only as 
$T\rightarrow\infty$. The parameters $a_i$ and $b_3$ are 
taken from Ref.~\cite{Rossner:2006xn}, where they were
determined to reproduce lattice data for the
thermodynamics of pure gauge lattice QCD up to about 
twice the critical temperature\footnote{At much higher temperatures, where 
transverse gluons begin to dominate, the 
PNJL model is not supposed to be applicable.}. 
The resulting parameters are 
\bea
a_0 = 3.51~,~~a_1 = -2.47~,~~a_2 = 15.22~,~~b_3 = -1.75~.
\nonumber
\eea
The critical temperature $T_0$ for deconfinement in the pure gauge sector is
fixed at 270 MeV in agreement with lattice results.

The NJL part of the model involves three parameters: the bare quark mass which 
we take equal for $u$- and $d$-quarks, the coupling strength $G$ and a 
three-momentum cutoff $\Lambda$. We take those from Ref.~\cite{Ratti:2005jh}:
\bea
m_{u,d} = 5.5~\mathrm{MeV}~,~~G = 10.1~\mathrm{GeV}^{-2}~,~~\Lambda = 0.65~
\mathrm{GeV}~,
\nonumber
\eea 
fixed to reproduce the pion mass and decay constant in vacuum and 
the chiral condensate as $m_\pi =$ 139.3 MeV, $f_\pi =$ 92.3 MeV and 
$\langle\bar{\psi}_u\psi_u\rangle = - (251$ MeV)$^3$.

After performing a bosonization of the PNJL action and introducing
scalar and pseudoscalar auxiliary fields, $\sigma$ and ${\vec\pi}$, the PNJL
thermodynamic potential becomes:
\bea
&&\!\!\!\!\!\!\!\!\Omega\left(T,\mu_q,\sigma,\Phi,\Phi^*\right)\!=\!\,\mathcal{U}
\left(\Phi,\Phi^*,T
\right)+\frac{\sigma^{2}}{2G}
\\
&&
-2N_f\int\frac{\mathrm{d}^3p}{\left(2\pi\right)^3}
\left\{T\!\,\ln\!\left[1+3\,{\Phi}\,\mathrm{e}^{-\left(E_p-\mu_q\right)/T}\!\!+
3\,{\Phi^*}\,\mathrm{e}^{-{2}\left(E_p-\mu_q\right)/T}+
\!\!\,\mathrm{e}^{-3\left(E_p-\mu_q\right)/T}\right]\right.
\nonumber\\
&&\!\!\!\!+\!\left.T\! \,\ln\!\left[1\!+\!\,3\,{\Phi^*}\,\mathrm{e}^{-\left(E_p+
\mu_q\right)/T}\!\!+\!\,
3\,{\Phi}\,\mathrm{e}^{-{2}\left(E_p+\mu_q\right)/T}\!\!+
\!\,\mathrm{e}^{-{3}\left(E_p+\mu_q\right)/T}\right]+3\,\Delta E_p\,\theta
\left(\Lambda^2-\vec{p}^{~2}\right)\right\}
\nonumber
\eea
where the quark quasiparticle energy is $E_p=\sqrt{\vec{p}^{\,\,2}+m^2}$ and the 
dynamical (constituent)
quark mass is the same as in the standard NJL model:
$m=m_0-\sigma=m_0-G\langle\bar{\psi}\psi\rangle$. The last term in the previous equation involves the difference $\Delta E_p$ between the quasiparticle energy $E_p$ and 
the energy of free fermions (quarks). It is understood that for three-momenta 
$|\vec{p}\,|$ above the cutoff $\Lambda$ where NJL interactions are 
``turned off", $\sigma$ is set to zero.

In general, the Euclidean action ${\cal S}_E$ is formally complex in the 
presence of the temporal gauge
field $\phi$~\cite{Simonthesis,Fukushima:2006uv}. It is real only at vanishing 
chemical potential, $\mu_q=0$. At nonvanishing chemical potential one has in 
addition $\langle\Phi\rangle\neq\langle\Phi^*\rangle$~\cite{dpz05}.
Fluctuations beyond mean field are at the origin of 
$\langle\Phi\rangle\neq\langle\Phi^*\rangle$ for $\mu_q\neq 0$~\cite{simon2}
(see the more detailed discussion in~\cite{Rossner:2006xn}).
This paper deals with self-consistent solutions and predictions of the 
mean-field equations
\beq
\frac{\partial\,\mathrm{Re}\,\Omega}{\partial\varphi}=0
\label{mf}
\eeq
with $\varphi = \sigma, \phi_3, \phi_8$. At this mean field level, the 
additional constraint of $\phi_i$ being real implies
that the action is minimized by $\phi_8 = 0$. It follows that 
$\Phi  \sim\mathrm{Tr}\exp(i\lambda_3\,\phi_3/T)$ is real 
(i.e. $\Phi =\Phi^*$ at the mean field level).
The temperature dependence of the Polyakov loop and chiral condensate, obtained
from Eq.~(\ref{mf}) can be found in Ref.~\cite{Rossner:2006xn}.

A Taylor expansion of the pressure in $\mu_q/T$ has been performed
in Ref.~\cite{Rossner:2006xn} in order to compare
with corresponding lattice data~\cite{Allton:2005gk}. The pressure is written 
as a series
\bea
\frac{p\left(T,\mu_q\right)}{T^4}&=&\sum_{n=0}^{\infty}c_n\left(T\right)
\left(\frac{\mu_q}{T}\right)^n;
\nonumber\\
c_n\left(T\right)
&=&\left.\frac{1}{n!}\frac{\partial^n\!\left( p\left(T,\mu_q\right)/T^4\right)}
{\partial\left(\mu_q/T\right)^n}\right |_{\mu_q=0}.
\label{pressure}
\eea
The Taylor expansion coefficients $c_2,~c_4$ and $c_6$, calculated in the
PNJL model, turn out to  agree very well 
with the corresponding lattice data. 

The PNJL model calculations can provide 
useful insights concerning the convergence of the expansion (\ref{pressure}).
The sign problem at non-zero chemical potential restricts lattice QCD
thermodynamics to extrapolations including only the first few terms of
Eq.~(\ref{pressure}), whereas the model is not restricted to such truncations.
In Ref.~\cite{Ratti:2006wg} we have shown that, for the pressure, 
very good agreement is reached between the full result and the truncated one, 
already
at fourth order in $\mu_q/T$ and up to $\mu_q/T\sim1$. In the case of the quark
number density, the agreement is still very good at small chemical potentials, 
but discrepancies are observed at large $\mu_q/T$ in the vicinity of the phase
transition.

\begin{figure}
\begin{center}
\begin{minipage}[t]{.48\textwidth}
\includegraphics*[height=168pt]{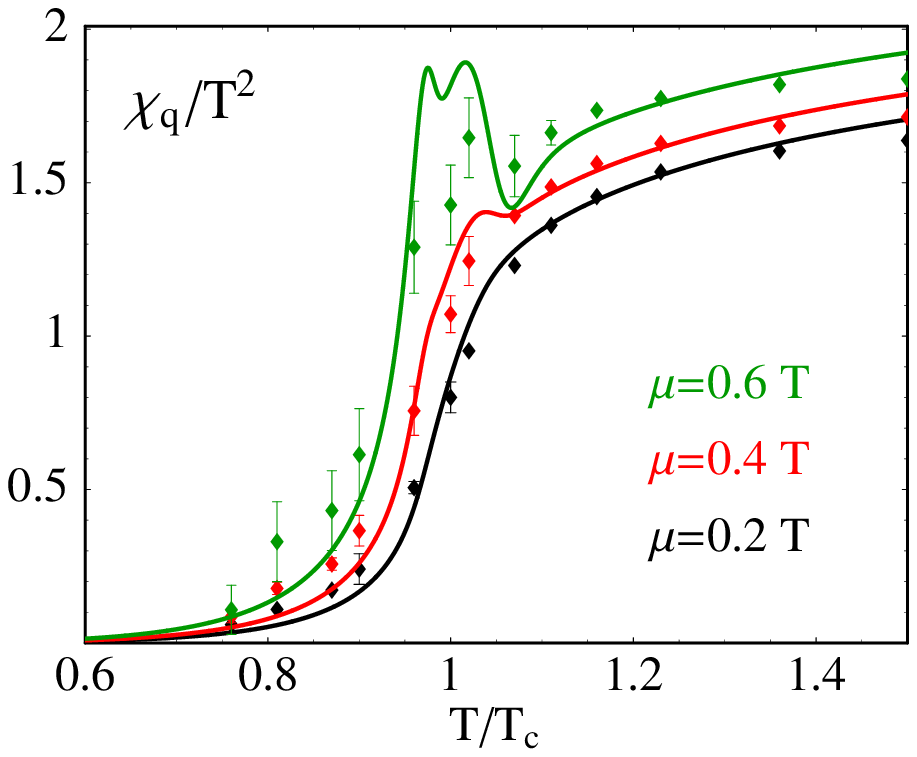}
\end{minipage}
\hfill
\begin{minipage}[t]{.48\textwidth}

\includegraphics*[height=165pt]{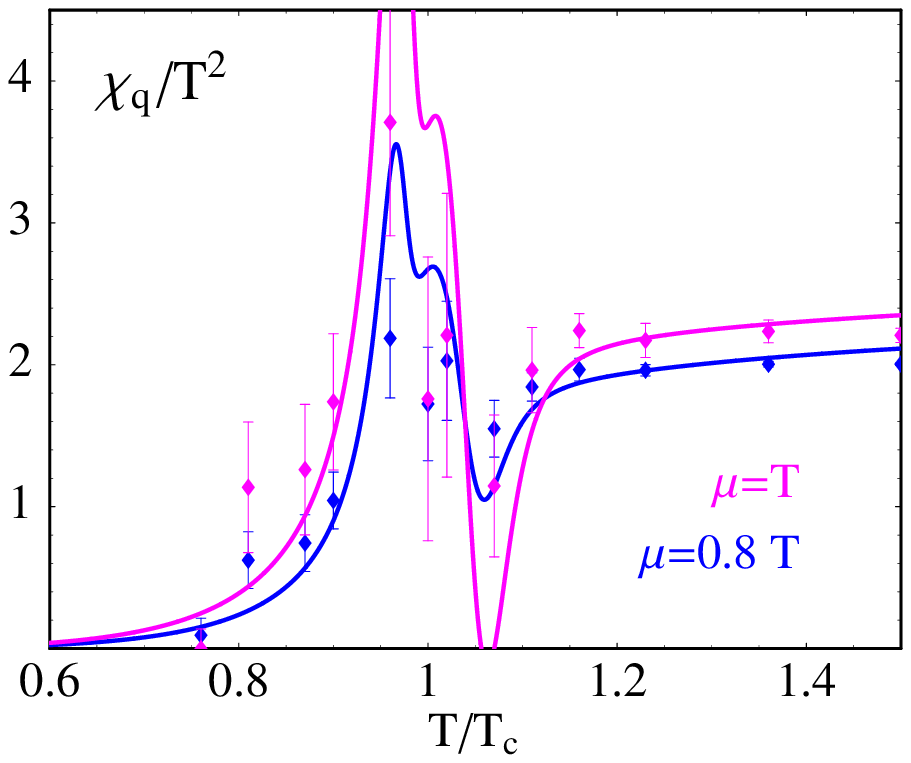}
\end{minipage}
\caption{
\footnotesize
Scaled quark number susceptibility $\chi_q$ as function of $T/T_c$ for 
different 
values of $\mu_q/T$. The curves are PNJL model results obtained from the Taylor
series expansion in $\mu_q/T$ around $\mu_q=0$ (see Eq.~(\ref{taylor})).
Corresponding lattice data are taken from Ref.~\cite{Allton:2005gk}.
\label{fig1}}
\end{center}
\end{figure}

In the present paper we investigate the quark number 
susceptibilities defined as
\bea
\frac{\chi_q\left(T,\mu_q\right)}{T^2}=\frac{\partial^2\!\left(p/T^4\right)}
{\partial\!\left(\mu_q/T\right)^2}~~.
\eea
The isospin-symmetric case is considered, with equal
chemical potential for up and down quarks.
The full result for $\chi_q$ is compared with the truncated one given by the
Taylor expansion
\bea
\frac{\chi_q\left(T,\mu_q\right)}{T^2}=2\,c_2+12\,c_4\left(\frac{\mu_q}{T}\right)^2
+30\,c_6\left(\frac{\mu_q}{T}\right)^4.
\label{taylor}
\eea
In Fig.~\ref{fig1} we show a comparison between truncated PNJL result obtained
from Eq.~(\ref{taylor}), and the corresponding lattice data, consistently 
expanded to the same order in $\mu_q/T$. One finds remarkably good agreement 
between the lattice data and the PNJL predictions. At large chemical potential
$\mu_q/T\sim1$ we observe, consistently with the lattice data, a 
pronounced peak sligthly below $T_c$, followed by a dip around
$T\sim1.05~T_c$. This feature was also discussed in \cite{Allton:2005gk}.

\begin{figure}
\begin{center}
\begin{minipage}[t]{.48\textwidth}
\includegraphics*[height=168pt]{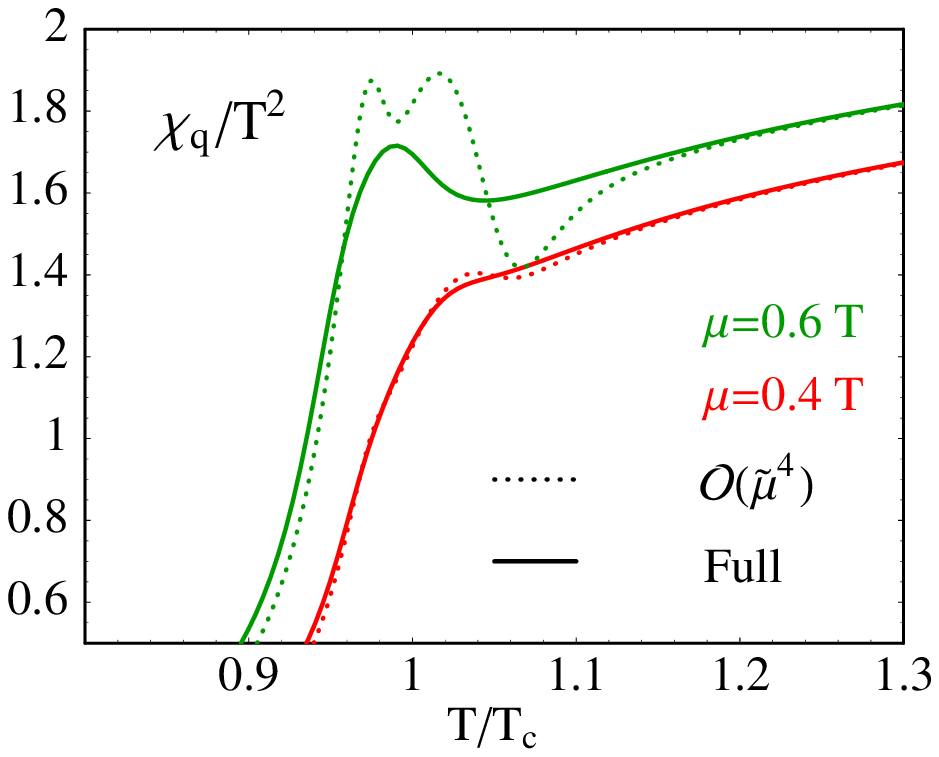}
\end{minipage}
\hfill
\begin{minipage}[t]{.48\textwidth}
\includegraphics*[height=165pt]{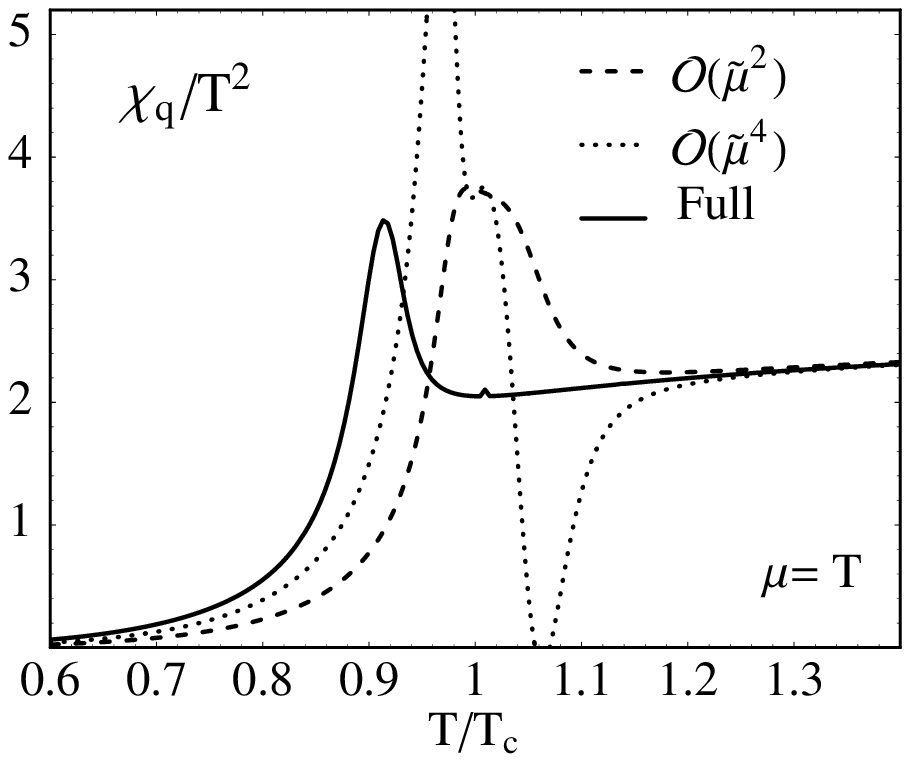}
\end{minipage}
\caption{
\footnotesize
Left: comparison between the full and truncated PNJL results
for two different (small) chemical potentials. Right:
comparison between full and truncated results, at different 
orders, for $\mu_q/T=1$ (see Eq.~(\ref{taylor})).
\label{fig2}}
\end{center}
\end{figure}

Fig.~\ref{fig2} presents a study of the convergence of the Taylor expansion in
${\tilde \mu}=\mu_q/T$. A comparison is made between full PNJL results
and those truncated at fourth and sixth order in ${\tilde \mu}$, for a sequence
of chemical potentials $0.4\leq\mu_q/T\leq1$. For low
chemical potentials, the agreement between the full result and the truncated
one is rather good. Only in the case of $\mu_q/T=0.6$ and in the region of $T$
around the phase transition, the first discrepancies start to appear.

The right panel of Fig.~\ref{fig2} shows a comparison between the full PNJL
result (continuous line), the one truncated at fourth order (dashed line) and 
the one truncated at sixth order (dotted line) 
for $\mu_q/T=1$. Evidently, the Taylor
series converges badly in the region around the phase transition. 
The full result shows a peak, not a divergence, 
around $T/T_c\sim0.9$, indicating that the critical temperature
for the phase transition decreases with increasing chemical potential.
The coefficient $c_6$ is sizeable around the phase transition,
so that the sixth order contribution differs significantly from that of
fourth order. The dip above $T_c$ occurring at sixth order is 
entirely an effect of the truncation of the series. It is absent in the
full result. It reflects the corresponding dip exhibited by $c_6$: in the
fourth order result (dashed line) it is in fact absent as well.  
The height of the peak in the full result is much reduced as compared to
the corresponding one found at sixth order. No divergence is observed in the
full result. In fact, our model predicts that the phase transition is a 
crossover, for this value of $\mu_q$. This analysis indicates that the 
truncated $\chi_q$ can be misleading in the sense that it incorrectly suggests
the possibility of a first-order transition which is not realized in the
full susceptibility.

From the present work we can draw the following conclusions.
The available lattice results for quark number susceptibilities at finite 
chemical potential, obtained from a Taylor
expansion in $\mu_q/T$ around $\mu_q=0$, can be successfully reproduced in the
framework of the PNJL model through an analogous Taylor expansion truncated
at the same order in $\mu_q/T$. The comparison between the full 
PNJL result at finite chemical potential and the truncated one shows good 
convergence of the Taylor series up to $\mu_q/T\leq0.4-0.6$. 
For larger values of $\mu_q/T$, significant discrepancies are observed between
full and truncated results in the region around the phase transition. The
singular behaviour of the susceptibilities observed in the sixth-order result
is not present in the full calculation. The transition 
predicted by our model for $\mu_q/T=1$ is still a crossover, reflected by the 
finite height of the peak in the corresponding quark number susceptibilities. 


\end{document}